\documentstyle[twocolumn,floats,psfig,aps]{revtex}

\begin{document}

\twocolumn[\hsize\textwidth%
\columnwidth\hsize\csname@twocolumnfalse\endcsname

\title{Persistent Current in 
an Artificial Quantum Dot Molecule}
\author{R. Kotlyar and S. Das Sarma}
\address{Department of Physics\\
University of Maryland \\
College Park, Maryland 20742-4111}
\date{October 8, 1996} 
\maketitle

\begin{abstract}
Using an exact diagonalization technique within a generalized
Mott-Hubbard Hamiltonian, we predict the existence of a ground state
persistent current in coherent two-dimensional semiconductor quantum dot
arrays pierced by an external magnetic flux. The calculated persistent
current, which arises from the nontrivial dependence of the ground state energy
on the external flux, exists in isolated arrays without any periodic boundary
 condition.
 The sensitivity of the calculated persistent current to interaction and
 disorder is shown to reflect the intricacies of various
 Anderson-Mott-Hubbard quantum phase transitions in two dimensional
systems.
\end{abstract}
\pacs{73.20.Dx, 73.40Gik, 71.30.+h, 05.30.Fk}
\vskip1pc]
\narrowtext

Significant recent advances in semiconductor materials growth and 
nanolithography techniques now allow\cite{exp} for the fabrication of 
multiple quantum dot systems in modulation-doped two dimensional 
electron structures which are quantum mechanically coherent, and 
may therefore be considered two dimensional ``artificial molecules'' 
(with individual quantum dots being the ``atomic'' constituents of this 
artificial molecule). 
 Theoretical work on multi-dot systems
has mostly concentrated on the two limiting 
situations\cite{kirczenow,middleton} (coherent dots with no Coulomb 
interaction\cite{kirczenow} and Coulomb blockade of individual 
dots\cite{middleton}). In this Letter we consider coherent 
two dimensional (2D) arrays of quantum dots (i.e. 2D artificial molecules) 
taking into account\cite{stafford94,klimeck} both quantum 
fluctuations arising from interdot hopping and electron-electron interaction 
effects through a generalized Mott-Hubbard Hamiltonian\cite{stafford94}.
Our main result is the prediction of an 
{\it{equilibrium persistent current}} in finite 2D arrays of semiconductor 
quantum dots in the presence of an applied magnetic field transverse to the 
2D plane. A measurement of this persistent current in small 2D arrays 
will be definitive evidence for the formation of an artificial quantum dot 
molecule. The predicted persistent current is a periodic function of 
the external flux and exhibits clear signature of the competition between 
coherence and interaction in artificial 2D molecules. We include disorder effects in our calculation and show that our predicted equilibrium 
persistent current should be experimentally observable in currently 
(or soon-to-be) accessible 2D quantum dot arrays.

It is well-known\cite{london,byers,buttiker} that gauge invariance 
and single-valuedness of electron wavefunction allow for the existence 
of a ground state persistent current in normal metal rings surrounding an 
external magnetic flux. The intrinsic magnetic moment associated with 
this persistent current, which is proportional to the current in 
one dimensional (1D) rings, is an oscillatory function of the external flux 
with a period equal to the elementary flux quantum $\phi_{0}=h/e$. 
The existence of such an oscillatory persistent current in 
normal metal rings has been 
experimentally verified\cite{levy,chand,mailly}. 
 Our predicted 
ground state persistent current in 2D quantum dot arrays has some 
significant differences with the existing persistent current theoretical 
analyses in the 1D ring geometry. The system we consider is a finite 
coherent 2D 
quantum dot array where the one electron states are in the atomic tight binding limit (i.e. the electron states tend to be strongly localized on 
individual quantum dots except for weak interdot quantum tunneling) 
in contrast to metallic systems which are nearly free electron like. 
In addition, we consider a finite 2D array without any periodic 
boundary condition in contrast to the 1D ring geometry, 
which, by definition, is periodic. Another significant aspect of our 
prediction is the inclusion of quantum tunneling, disorder, and Coulomb 
interaction effects on equal footings because in real semiconductor 
quantum dot arrays all of these effects are expected to be important.

We model the finite square 2D quantum dot array of size 
$L_{x} \times L_{y} = L$ dots through the following physically 
motivated generalized 2D Mott-Hubbard type strongly 
correlated Hamiltonian\cite{stafford94} 
($c, \ c^{\dag}$ are electron field operators and $\hat{n}$ is the 
density operator):
\begin{eqnarray}
H=\sum_{i,\alpha} \varepsilon_{\alpha} {{c}^{\dag}}_{i\alpha} c_{i\alpha}+\frac{U}{2} \sum_{i} {{\hat{n}}_{i}}^{2}+
\sum_{ij} \frac{V_{ij}}{2} {\hat{n}}_{i} {\hat{n}}_{j} \nonumber \\ 
- \sum_{<i,j>, \alpha} (t_{\alpha, ij}e^{i\phi_{ij}} {{c}^{\dag}}_{i\alpha} c_{j\alpha} + h.c.),
\label{1}
\end{eqnarray}
where $\phi_{ij}= \frac{e}{\hbar } \int_{ij} \vec{A} \cdot \vec{l_{ij}}$ 
is the usual Peierls gauge phase factor arising from electron hopping 
on a lattice in a transverse external magnetic flux (with $\vec{A}$ 
the vector potential and $\vec{l_{ij}}$ the spatial vector along a 
plaquette). The indices $i, \ j$ are the quantum dot spatial positions 
in the array and $\alpha$ denotes an electron state on the quantum dot. 
The generalized Mott-Hubbard parameters $U, \ V$ and $t$ have the usual 
significance of the on-site Coulomb interaction, the long range part 
of the Coulomb interaction and the hopping kinetic energy term 
respectively within the tight binding description of the 2D array. 
We consider two spin-split energy 
levels $\varepsilon_{\alpha}$ per dot due to quantum 
confinement in a dot.  
 We assume the magnetic field 
to be weak enough for single-particle level crossing effects 
not to be important.
 The  Coulomb 
interaction in the array is expressed in terms of the capacitance matrix 
representation\cite{middleton}. 
The quantum dot array is in 
the tight binding atomic  
limit, and we therefore keep only the nearest neighbor tunneling 
term in $t_{ij} \equiv t$. Within the model defined by Eq. (1), 
$U, \ V, \ t$ are parameters to be given as inputs using a physical 
description of the quantum dot array. 

We are interested in calculating 
the equilibrium persistent current flowing in the array at $T=0$ in the 
presence of an external flux threading the 2D lattice. It is easy to show by 
using commutation properties of $H$ with the polarization 
operator that the transverse component of the magnetic moment 
density operator, 
$m_{z}$, is given by the gauge invariant result 
\begin{equation}
\label{2}
m_{z} = - \frac{\partial H}{\partial \Phi},
\end{equation}
where $\Phi$ is the total uniform magnetic flux piercing
the finite 2D system.
(We note that the external magnetic field is in the $z$ direction and with the 
2D array in the $x-y$ plane, and therefore $m_{z}$ is the only non-zero 
component of the  magnetization density.) A non-zero 
value of the static magnetization density, as measured by a finite $m_{z}$, 
is equivalent to the existence of a ground state persistent current (whose 
strength is exactly the same as the measured value of $m_{z}$ because of the 
two dimensional nature of the problem). Our goal 
is to calculate the equilibrium magnetic moment density 
for 2D quantum dot 
arrays using Eqs. (1) and (2). 
 An experimental measurement of a finite
$m_{z}$ (as a function of external flux) will be a direct verification
of the quantum dot molecule formation. 

To obtain the equilibrium persistent current,
$I \equiv m_{z}$,  
we calculate the ground state energy by doing an 
{\it{exact diagonalization}} of $H$, defined by Eq. (1), in the 
basis of the total 
number of electrons $N$ in the array and the total  
spin $S_{z}$ in the direction of the field, and then 
perform minimization over $S_{z}$ to find the stable 
ground state. The magnetic 
moment density in a finite 2D quantum dot array
is calculated using  Eq. (2). 
 Existing semiconductor  nanofabrication techniques are {\it{at best}}
 capable of producing coherent arrays which are $2 \times 2$ to 
$3 \times 3$ quantum dots in size. The Lanczos diagonalization 
technique that we use is capable of giving the exact interacting 
ground state for upto a $3 \times 3$ array of our interest.
 We calculate the persistent current in 
{\it{interacting}} 2D arrays upto $L=3 \times 3$ sizes 
for all band fillings $n(=N/2L)$ by varying the number of electrons 
$N$ in each system between $N=1$ and $N=L$. We use 
 realistic Mott-Hubbard 
parameters which approximately correspond to 2D GaAs quantum dot systems 
(at $T=0$). We choose\cite{stafford94} the intradot single particle level 
spacing $\Delta=0.3 \ meV$  to be comparable in magnitude to the 
interdot hopping parameter $t=0.1 \ meV$, with 
the intradot Coulomb charging energy $U \sim 1 \ meV$. 
With this choice of parameters the 2D quantum dot array is in the interesting 
coherent ``molecular'' state referred to as the ``Collective Coulomb 
Blockade'' regime\cite{stafford94}, where both coherence and 
interaction play comparable non-perturbative 
roles, and the Coulomb blockade of individual dots is destroyed. 
Disorder is included in our calculation through a spin-independent 
parameter $W \sim \Delta$ which denotes the half-width of a uniform 
distribution of random on-site quantum dot energies centered around $\Delta$. 
The spin splitting (set to be $0.03\Delta$) 
of the single particle energy levels is assumed to be independent 
of $W$.
\begin{figure}[h]
\label{fig1}
\psfig{figure=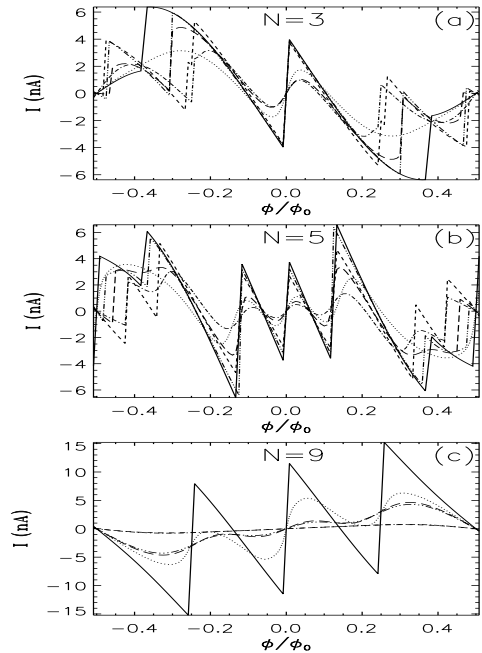,width=0.9\columnwidth}
\caption{
Typical calculated persistent current
for three different values of band filling
($n=1/6 (a), \ 5/18 (b), \ 1/2 (c) $) in the $3 \times 3$ array is
plotted versus the magnetic flux
through the elementary cell in units of $\phi_{0}$
for six different sets of $U,\ V, \ W$
(see the legend in the Fig. 2 for the key to each line in the plot)
in the $3 \times 3$ system.
}
\end{figure}

We show our typical calculated persistent current as a function of the 
external flux in Fig. 1 for three different values of band filling 
($n=1/6, \ 5/18, \ 1/2$) in the $3 \times 3$ system. Results for 
six different situations (as described in the figure captions) for various 
values $U, \ V,$ and $W$ are shown for each filling with $t$, $\Delta$ 
fixed throughout the calculation. The flux $\phi$ is through each 
elementary cell (or plaquette) of the $3 \times 3$ system, and the 
calculated equilibrium persistent current is in absolute units. 
The most 
significant feature of Fig. 1 is the existence of a persistent current, 
which is 
periodic in $\phi/\phi_{0}$ 
({\it{note}} that the total flux through the system is 
$\Phi=4\phi$ by virtue
of the four equivalent plaquettes in the $3 \times 3$ quantum dot array), 
whose magnitude is reasonably large ($\geq$ nA). The magnitude of the current depends in a rather complicated way on the Mott-Hubbard parameters 
$U, \ V$  and $W$, and also on the band-filling (in particular, 
on whether $n=1/2$ or not). We can, however, make the following 
qualitative observations: (1) Away from half-filling, finite disorder is the most prominent destructive effect on the current amplitude, with finite 
interaction ($U,\ V \ \neq 0$) producing some enhancement of the 
disordered current but never to the free electron ($U, \ V, \ W \ = \ 0$) 
value. Without any disorder ($W=0$), finite $U$ and $V$ have 
reasonably 
small 
effects on the current magnitude away from half-filling. 
(2) At half-filling, however, finite on-site 
Coulomb repulsion ($U \neq 0$) by itself dramatically suppresses 
the persistent current with the long-range Coulomb interaction ($V\neq 0$) 
opposing this dramatic suppression due to finite $U$. 
(3) Disorder, by itself ($W \neq 0, \ U, \ V \ =0$), 
seems to produce similar quantitative suppression 
of the persistent  current at all fillings 
except at half-filling where disorder enhances (slightly) 
the interacting clean system current. 
(4) Long range Coulomb repulsion ($V \neq 0$) always  
 enhances the current magnitude from its finite $U$ and $W$ values. 
(5) The various structures (jumps and discontinuities) in the current arise 
from 
many-body energy level crossings
 which occur in the system at specific fillings and values 
of the magnetic flux. The discontinuities of the persistent 
current  seen in Fig. 1 correspond to the change of the total orbital 
momentum in the 2D array, which can be accompanied by  
spin-flip transitions in the system. Finite disorder
 ($W \neq 0$) smoothens the discontinuities 
in the noninteracting ($U, \ V \ =0$) current, and 
removes some of the discontinuities in the 
interacting ($U, \ V \ \neq 0$) current. There are 
no discontinuities or spin flips at half-filling
in the interacting system (Fig. 1(c)).
(6) The sign of the persistent current at zero magnetic 
flux is generally dependent on the electron 
filling and the geometry of the array. We find 
the interacting ($U \neq 0, \ V, \ W \ =0$)
persistent current in $2 \times 2$, $3 \times 2$, $4 \times 2$, 
and $3 \times 3$ arrays to be diamagnetic for 
$n=1/2L$ and paramagnetic for $n=1/2$. The random 
disorder of strength $W$ ($=\Delta$, as used in our 
calculations) does not 
change the sign of the persistent current. 
(7) The current distribution in the clean 2D array is 
determined by the spatial symmetries of the system 
and band filling in the array. In the clean 
$3 \times 3$ lattice the current flows along the 
boundary for all electron fillings in the system. 
However, the persistent current distribution is 
disorder realization dependent.

The most important feature of our results, the strong 
suppression of the persistent current at half-filling by the on-site 
Coulomb repulsion $U$, is a direct {\it{finite size}} 
manifestation of the Mott-Hubbard metal-insulator transition, which is 
a true phase transition in the thermodynamic limit. The long-range 
Coulomb repulsion ($V \neq 0$) opposes this transition in a finite system 
(leading to an enhancement in the finite $U$-suppressed current 
amplitude for $V \neq 0$ at half-filling). 
Effects of finite $V$ and $W$ on the Mott-Hubbard transition 
in a 2D Hubbard model (in the thermodynamic limit) are 
not rigorously known, but our results are consistent with the expectation 
that, in general, finite values of $V$ and $W$ should oppose the 
metal-insulator 
Mott transition at half-filling. 
\begin{figure}[h]
\label{fig2}
\psfig{figure=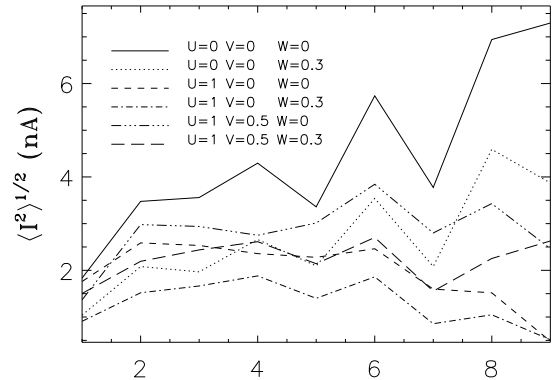,width=0.9\columnwidth}
\caption{
A plot of the RMS persistent current $\langle I^{2} \rangle^{1/2}$
 versus the number of
electrons $N$ in the $3 \times
 3$ array for the six different sets of $U,\ V, \ W$
(as shown in the legends)
in
the $3 \times 3$ system.
 All energies are given in the $meV$ units.
}
\end{figure}
To further quantify the relative 
magnitudes of the effects of finite $U, \ V, \ W$ 
on the persistent current at various band fillings, we show in Fig. 2 our 
calculated root mean square current amplitude as a function of the 
number of electrons  in the $3 \times 3$ quantum dot array for 
various values of Mott-Hubbard parameters. The rms current magnitude,
 $\langle I^{2} \rangle^{1/2}$, is obtained by averaging the typical current 
(Fig. 1) over one flux period. The six different parameter sets with different 
values of $U, \ V, \ W$ in Fig. 2 clearly show the dominant effect of the 
Mott-Hubbard transition (for finite $U$) at half-filling ($N=9$), the generic 
destructive effect of finite disorder at all fillings, and the generic 
tendency of the long-range Coulomb interaction 
homogenize the electron density in the system\cite{giamarchi} and 
consequently
to enhance the persistent 
current. 
\begin{figure}[h]
\label{fig3}
\psfig{figure=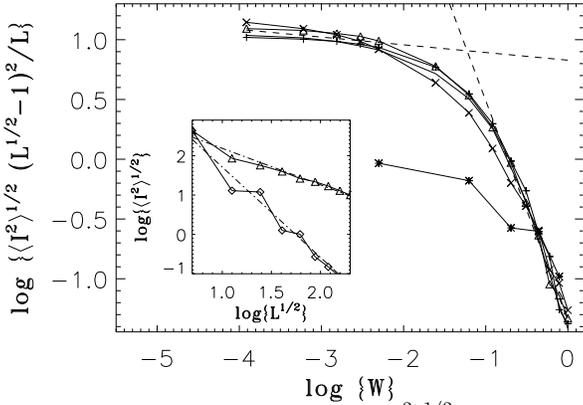,width=0.9\columnwidth}
\caption{
A log-log plot of $\langle I^{2} \rangle^{1/2}$
versus $W$
averaged over 100 disorder realizations
for each value of $W$ is shown
 ($U\ = \ V\ = \ 0$)
at half-filling in
 $3 \times 3$ (crosses), $4 \times 4$
(triangles), $5 \times 5$ (solid line), $6 \times 6$
(pluses).
Asterisks show the interacting ($U=1\ meV, V=0$) results
at half-filling for the $3 \times 3$ system.
Inset:
$\langle I^{2} \rangle^{1/2}$
at half-filling
versus system sizes ($L$): triangles (2D array), squares (1D ring).
 (The average is taken over the range of the
same total flux
through the systems, which varies from $-\phi_{0} / 2$ to $\phi_{0} / 2$.)
 The dashed lines in the plot are the best linear fits to
the corresponding data as described in the text.
}
\end{figure}

In the well-studied 
single 1D ring geometry, the effect of disorder (in the absence 
of any Coulomb interaction) on the persistent current is an exponentially 
strong suppression\cite{cheung} of the current arising from the 
Anderson localization phenomenon associated with all electronic states 
being exponentially localized in one dimension in the presence of any 
finite disorder. 
For our 2D quantum dot array the effect of finite disorder 
($W \neq 0$) is subtle (and much softer than in one dimension) 
because localization effects are logarithmically 
weak\cite{belitz} in two dimensions 
for small $W$. (In fact, in the presence of an external magnetic 
field breaking the time reversal symmetry, the localization effect is even 
weaker\cite{belitz} 
and may be hard to discern in a finite sample.) This explains 
the rather benign effect of having a finite $W$ in Figs. 1 and 2 of our 
results, which should be contrasted with 1D ring 
calculations where the persistent current is strongly suppressed by 
finite values of $W$. To better quantify our finite disorder results we show 
in Fig. 3 a log-log plot at half-filling of the rms 
current $\langle I^{2} \rangle^{1/2}$ without any 
interaction effects ($U, \ V=0$), averaged 
over 100 disorder realizations for each value of $W$, as a function 
of the disorder strength $W$ for various system sizes 
($3 \times 3, \ 4 \times 4,\ 5 \times 5,\  6 \times 6$). In plotting these 
results, we have factored out a scale factor, 
$n_{c}=(L^{1/2}-1)^{2}/L$, so that the results for various system sizes 
fall on top of each other, showing approximate current scaling with 
system size and disorder. (The scale factor $n_{c}=(L^{1/2}-1)^{2}/L$ 
is the number of unit cells or plaquettes in each square array of size $L$.)
All four non-interacting disordered results scale very well 
whereas the one disordered result 
{\it{with}} interaction deviates from the universal scaling 
at low disorder strength where the Mott metal-insulator 
transition effect dominates. The two dashed straight lines in Fig. 3 give the 
best fits to weak and strong disorder scaled currents, leading to the following 
empirical results for the effect of disorder on the persistent current:  
$\langle I^{2} \rangle^{\frac{1}{2}} \  \ = \ 
 \left(L/n_{c} \right) \ g(W)$,   
with the scaling function $g(W)$ being given by
\begin{eqnarray}
\nonumber
 g(W) \sim \left\{ 
\begin{array}{ll}
W^{-\gamma}, \ \gamma=(6.4 \pm 2.8) \times 10^{-2},\ \ \  W < 1.55\  \pi t \\
W^{-\beta}, \ \beta=1.84 \pm 0.49, \ \ \ \ \ \ W > 1.55 \  \pi t.
\end{array}
\right.
\end{eqnarray}
The very small value of the scaling exponent $\gamma$ in the weak disorder limit is consistent with the expected 
logarithmic weak localization\cite{belitz} 
in a 2D array, which eventually crosses 
over to strong localization for large $W$. We also expect interaction effects 
to be particularly important at weak disorder, but not so at high disorder, 
which is what is seen in Fig. 3 even at half-filling. 
Weak disorder, in fact, produces anti-localization effect 
(Fig. 1(c) and 2) at half-filling 
{\it{in the presence of finite $U$}}, by 
slightly enhancing the persistent current from its 
finite $U$-suppressed value.
Whether 
these last two findings 
 remain true in the thermodynamic 
limit is, however, unknown.

Note that the scale factor $n_{c}$ in Fig. 3 indicates that the persistent 
current scales with the perimeter size of the array or the 
square root of the number of dots, $L^{1/2}$, in the system. We have 
explicitly verified this system size dependence by doing calculations for various system sizes at various fillings. As an example, we show as an inset 
of Fig. 3 a comparison of the system size dependence of the rms current 
in 1D ring and 2D array geometries, finding an approximately linear dependence on $L^{\alpha}$ with the best fit 
$\alpha \approx 0.46\ (1.1)$ in the 
2D array (1D ring). Thus, the persistent current is proportional 
to the length of the boundary in both 1D ring and 2D array 
geometry. 

In conclusion, we predict, based on an exact 
diagonalizaton study of the generalized disordered 
Mott-Hubbard Hamiltonian, 
 the existence of a ground state 
persistent current in finite 2D quantum dot arrays which should be 
experimentally observable. Direct observation of an 
equilibrium persistent current in coherent 2D quantum dot 
arrays will not 
only verify the formation of an artificial quantum dot molecule but  
will also shed light on the interplay among coherence, disorder, 
long- and short- range Coulomb interaction effects on 
Anderson-Mott-Hubbard quantum phase transitions.

The authors thank Charles Stafford for useful discussions during the 
early part of this work. This work was supported by the U.S. - O.N.R.


\begin{thebibliography}{10}
\bibitem{exp} F. R. Waugh {\it et al.}, Phys. Rev. B {\bf 53}, 1413 (1996); 
R. H. Blick {\it et al.}, Phys. Rev. B {\bf 53}, 7899 (1996); 
F. Hofmann {\it et al.}, Phys. Rev. B {\bf 51}, 13872 (1995); 
N. C. van der Vaart {\it et al.}, Phys. Rev. Lett. {\bf 74}, 4702 (1995); 
D. Dixon {\it et al.}, Phys. Rev. B {\bf 53}, 12625 (1996). 
\bibitem{kirczenow}
G. Kirczenow, Phys. Rev. B {\bf 46}, 1439 (1992). 
\bibitem{middleton}
A. A. Middleton and N. S. Wingreen, Phys. Rev. Lett. {\bf 71}, 3198 (1993).
\bibitem{stafford94}
C. A. Stafford and S. Das Sarma, Phys. Rev. Lett {\bf 72}, 3590 (1994).
\bibitem{klimeck}
G. Klimeck, G. Chen, and S. Datta, Phys. Rev. B {\bf 50}, 2316 (1994); 
B {\bf 50}, 8035 (1994).
\bibitem{london}
F. London, J. Phys. Radium {\bf 8}, 347 (1937). 
 \bibitem{byers}
N. Byers and C. N. Yang, Phys. Rev. Lett {\bf 7}, 46 (1961).
\bibitem{buttiker} 
M. Buttiker, Y. Imry and R. Landauer, Phys. Lett. {\bf  96A}, 365 (1983).
\bibitem{levy}
L. P. Levy {\it et al.}, Phys. Rev. Lett {\bf 64}, 2074 (1990). 
\bibitem{chand}
V. Chandrasekhar,  {\it et al.}, Phys. Rev. Lett {\bf 67}, 3578 
 (1991). 
\bibitem{mailly}
D. Mailly  {\it et al.}, Phys. Rev. Lett {\bf 70}, 2020
 (1993). 
\bibitem{giamarchi}
T. Giamarchi and B. S. Shastry, Phys. Rev. B {\bf 51}, 10915 (1995) 
and references therein.
\bibitem{cheung}
Ho-Fai Cheung {\it et al.}, Phys. Rev. B {\bf 37}, 6050
 (1988).
\bibitem{belitz}
D. Belitz and T. Kirkpatrick, Rev. Mod. Phys. {\bf 66}, 261 (1994). 
\end{thebibliography}
\end{document}